\documentclass{aa}

\usepackage{natbib}
\bibpunct{(}{)}{;}{a}{}{,} 
\usepackage{graphicx}
\usepackage[varg]{txfonts}
\usepackage[normalem]{ulem}
\usepackage{amsmath, mathtools, esdiff}
\usepackage{xspace}

\usepackage[print-unity-mantissa = false]{siunitx}
\usepackage{xcolor}

\newcommand{\dint}[1]{~\mathrm{d}#1}

\DeclareSIUnit[]{\kb}{k_B}
\DeclareSIUnit[]{\yr}{yr}
\DeclareSIUnit[]{\Gyr}{\giga\yr}
\DeclareSIUnit{\bary}{m_u}
\DeclareSIUnit{\MJ}{M_J}
\DeclareSIUnit{\ME}{M_E}
\DeclareSIUnit{\RJ}{R_J}
\DeclareSIUnit{\RJeq}{R_J^{eq}}
\DeclareSIUnit{\kbperbary}{\kb \per \bary}
\DeclareSIUnit{\erg}{erg}
\DeclareSIUnit{\bar}{bar}

\newcommand{\Zcore}{Z_\mathrm{core}^{\dagger}}
\newcommand{\Zenv}{Z_\mathrm{env}^{\dagger}}
\newcommand{\mmid}{m_\mathrm{mid}}

\newcommand{\sRefCore}{s^{\odot}_\mathrm{core}}
\newcommand{\sRefEnv}{s^{\odot}_\mathrm{env}}

\newcommand{\sCore}{s^{\dagger}_\mathrm{core}}
\newcommand{\sEnv}{s^{\dagger}_\mathrm{env}}

\newcommand{\mesa}{\texttt{MESA}\xspace}

\newcommand{\nsims}{4,250\xspace}
\newcommand{\tSim}{\SI{5}{\Gyr}\xspace}
\newcommand{\dHist}{\qty{10}{\percent}\xspace}
\newcommand{\dmin}{\qty{1}{\percent}\xspace}

\newcommand{\ZCoreApprox}{0.4\xspace}

\newcommand{\ZEnvApprox}{0.05\xspace}

\newcommand{\sRefCoreBestFit}{\SI{9.29(0.47:0.72)}{\kbperbary}\xspace}
\newcommand{\sRefEnvBestFit}{\SI{10.40(0.47:0.48)}{\kbperbary}\xspace}

\newcommand{\sCoreApprox}{\SI{5}{\kbperbary}\xspace}
\newcommand{\sCoreBestFit}{\SI{4.98(3.00:2.57)}{\kbperbary}\xspace}

\newcommand{\sEnvApprox}{\SI{9.3}{\kbperbary}\xspace}
\newcommand{\sEnvBestFit}{\SI{9.32(0.48:0.58)}{\kbperbary}\xspace}

\newcommand{\RPrimBestFit}{\SI{1.89(0.40:0.49)}{\RJ}\xspace}

\begin{document}

\title{Further constraints on Jupiter's primordial structure}

\author{
H.~Knierim \inst{\ref{inst1}} \and
K.~Batygin \inst{\ref{inst2}} \and
R.~Helled \inst{\ref{inst1}} \and
L.~Morf \inst{\ref{inst1}} \and
F.~C.~Adams \inst{\ref{inst3}, \ref{inst4}}
}
\institute{
Department of Astrophysics, University of Zurich, Winterthurerstrasse 190, CH-8057 Zurich, Switzerland
\email{henrik.knierim@uzh.ch}\label{inst1}
\and
Division of Geological and Planetary Sciences, California Institute of Technology, Pasadena, CA, USA\label{inst2}
\and
Physics Department, University of Michigan, Ann Arbor, MI, USA\label{inst3}
\and
Astronomy Department, University of Michigan, Ann Arbor, MI, USA\label{inst4}
}
\date{Received 25 August 2025 / Accepted 1 December 2025}
\abstract{
The primordial structure of Jupiter remains uncertain, yet it holds vital clues on the planet's formation and early evolution.
Recent work used dynamical constraints from Jupiter's inner moons to determine its primordial state, thereby providing a novel, formation-era anchor point for interior modeling.
Building on this approach, we combine these dynamical constraints with thermal evolution simulations to investigate which primordial structures are consistent with present-day Jupiter.
We present \nsims evolutionary models of the planetary structure, including compositional mixing and helium phase separation, spanning a broad range of initial entropies and composition profiles.
We find that Jupiter's present-day structure is best explained by a warm (\sCoreBestFit), metal-rich dilute core inherited from formation. To simultaneously satisfy constraints on Jupiter's primordial spin, however, its envelope must have been significantly warmer (\sEnvBestFit) at the time of disk dispersal.
We determine Jupiter's primordial radius to be \RPrimBestFit.
These results provide new constraints on Jupiter's formation, suggesting that most heavy elements were accreted early during runaway gas accretion, and placing bounds on the energy dissipated during the accretion shock.
}
\maketitle
\section{Introduction} \label{sec:intro}
With its immense mass, Jupiter has exerted a profound influence on both the architecture and chemical evolution of the young Solar System \citep[e.g.,][]{Batygin_2015, Kleine_2020}.
It also serves as the archetype of a gas giant---a hydrogen-helium-dominated planet several hundred times the mass of Earth, and provides a crucial reference point for interpreting the growing population of giant exoplanets.
\par
Given that planetary interiors link planetary formation and present-day structure \citep[e.g.,][]{Helled2022_Jupiter, Knierim_2025}, Jupiter's interior has been studied intensively for decades.
Once considered a cold, homogeneous sphere of hydrogen and helium (H-He) \citep{Zapolsky_1969}, data from the Pioneer and Voyager missions established the classical three-layer model of a distinct, compact core surrounded by a deep H-He envelope \citep[e.g.,][]{Stevenson1982}. This framework became the standard for decades, refined with improved equations of state and atmospheric constraints from the Galileo mission \citep[e.g.,][]{Guillot2005}. It thus came as a genuine surprise when the Juno mission \citep[e.g.,][]{Bolton2017, Wahl2017} revealed a far more complex interior.
Among the most prominent features are helium sedimentation and a dilute core, first proposed by \citet{Stevenson_1977b} and \citet{Stevenson1985}, respectively.
These discoveries have spurred the development of a new generation of interior models \citep[e.g.,][]{ni2019,Nettelmann2025}, alongside advances in the high-pressure physics of hydrogen and helium \citep[e.g.,][]{Schoettler2018, Cozza_2025}.
\par
Despite these advances in characterizing Jupiter's current interior structure, its primordial configuration remains largely unknown.
Following the core accretion scenario \citep{Pollack1996}, recent studies have investigated Jupiter's formation \citep[e.g.,][]{Lozovsky2017, Helled2017, Stevenson2022}.
However, reconciling these primordial structures with Jupiter's present-day configuration, specifically its fuzzy core, remains challenging \citep{Vazan2018, Mueller2020, Meier2025}.
Using their new evolution code \texttt{APPLE} \cite{Sur_2024}, \cite{Tejada_Arevalo2025} and \citet{Sur2025, Sur2025b} recently identified initial conditions that can reproduce Jupiter's present-day structure. However, these studies made no statement about the origin of these primordial structures.
Recently, \citet{Batygin2025} (BA2025 hereafter) provided a constraint on Jupiter's primordial radius (given the moment of inertia, MoI) from dynamical constraints of the Jovian moon system.
In BA2025, the authors used these constraints to estimate Jupiter's primordial entropy, finding a specific entropy of $10.6$--\SI{11}{\kbperbary} and a primordial radius of $2.02$--\SI{2.59}{\RJ}, consistent with a ``warm-start'' scenario. However, their interior modeling was rather simplistic, assuming a non-evolving core-envelope structure. In addition, they did not attempt to match the properties of Jupiter at present-day.
In this study, we leverage this new primordial constraint to investigate initial models that not only reproduce Jupiter's present-day structure but also satisfy the first empirical constraint on its primordial state.
\par
This paper is structured as follows. In Sect. \ref{sec:methods}, we describe the numerical methods as well as the primordial constraint from BA2025. Section \ref{sec:results} shows how this new result can constrain Jupiter's primordial structure. These results are discussed and summarized in Sect. \ref{sec:discussion} and Sect. \ref{sec:conclusions}, respectively.
\section{Methods} \label{sec:methods}
Our numerical setup builds on the stellar evolution code \mesa \citep[][]{Paxton_2011, Paxton_2013, Paxton_2015, Paxton_2018, Paxton2019, Jermyn_2023}, which was extended to model planetary interiors with appropriate equations of state and opacities \citep[e.g.,][]{Mueller2020,Knierim_2025}. 
For hydrogen and helium we use the EOS of \cite{Chabrier_2021} while the heavy elements are represented by a 50-50 mixture of water (H$_2$O) and rock (SiO$_2$) \cite[e.g.,][]{Vazan2013}. 
The atmospheric boundary is the semi-gray atmosphere model of \citet{Guillot2010}, adjusting the visible opacity to match the Voyager radio occultation observations \citep{Gupta_2022}. 
Recently, we improved the stability and accuracy of convective mixing \citep{Knierim_2024} and included helium rain using the phase diagram of \citet{Schoettler2018} \citep[see][for details]{Knierim_2025}.
We also include uniform rotation, assuming conservation of angular momentum during the planet's evolution (for details, see Appendix~\ref{sec:rotation}).
Further information on the use of \mesa for giant planet modeling can be found in Helled et al.~(2025, submitted).
\par
To explore a wide range of initial structures, we generate a large number (\nsims) of random primordial composition and entropy profiles (for details, see Appendix \ref{sec:random_model_generation}), which we evolve for \tSim.
We identified models that reproduce Jupiter's present-day structure at some point during their evolution. Such a model can always be shifted to match Jupiter's age.
To assess how well a simulation matches present-day Jupiter, we minimize the relative deviation of the average radius, rotation rate, and two largest non-trivial gravitational moments ($J_2$ and $J_4$).
Given the radius-density profile of a simulation, we employ the Theory of Figures (ToF) \citep{Zharkov_1978} to calculate the resulting gravitational moments. Specifically, we use the ToF implementation presented in \citet{Morf_2024}, who developed the ToF to 10th order and improved its accuracy \citep{PyToF}.
\par
In addition, we examine the primordial constraint introduced by BA2025. Jupiter hosts four small inner moons interior to Io, two of which---Amalthea and Thebe---are thought to occupy primordial orbits. Notably, both exhibit small but non-zero inclinations of \ang{0.36} and \ang{1.09}, respectively.
These inclinations probably arose from second-order inclination resonances with Io, which implies that Io migrated outward due to tides after the dispersal of Jupiter's circumplanetary disk (CPD). Based on these interactions, \citet{Hamilton_2001} estimated that Io must have begun its outward migration at a semi-major axis $a_\mathrm{Io}^\dagger$ in the range $\xi = a_\mathrm{Io}^\dagger/\si{\RJ} = 4.11-5.09$. Here, the superscript $\dagger$ denotes the time at which the Jovian CPD dispersed, and we nondimensionalize by Jupiter's average radius (compared to the equatorial radius in BA2025).
For Io to stall at $a_\mathrm{Io}^\dagger$ during its inward migration when the CPD was still present, a mechanism must have reversed the disk torque---most likely the steep surface density gradient near the CPD's truncation radius.
In BA2025, the authors demonstrated that such a stall occurs when the dimensionless ratio $\zeta = a_\mathrm{Io}^\dagger / R_t$ is close to 1.13, where $R_t$ denotes the CPD's truncation radius.
This truncation radius is set by Jupiter's magnetic field and is therefore tied to its rotation rate and size \citep{Batygin_2018}.
By invoking angular momentum conservation, BA2025 derived: 
\begin{align}\label{eq:BA_2025}
    R_\mathrm{J}^{\dagger}=\left(\frac{I_\mathrm{J}}{I_\mathrm{J}^{\dagger}} \frac{\Omega_\mathrm{J}}{\chi \Omega_{\mathrm{br}}} \sqrt{\frac{\xi^{3}}{\zeta^{3}}}\right)^{1 / 2} \si{\RJ},
\end{align}
where $I_\mathrm{J}$ is Jupiter's normalized moment of inertia (NMoI), $\Omega_\mathrm{br} = \sqrt{G M_\mathrm{J}/R_\mathrm{J}^3}$ its present-day breakup rotation rate, and $\chi$ a constant approximately given by $\chi \approx 0.88$. 
Because Jupiter's present-day properties are well constrained, Eq.~\eqref{eq:BA_2025} provides a direct link between its primordial moment of inertia and radius. Hereafter, any reference to the BA2025 criterion or constraint refers to Eq.~\eqref{eq:BA_2025}.
\section{Results} \label{sec:results}
\begin{figure*}[ht!]
\centering
\includegraphics[width=\textwidth]{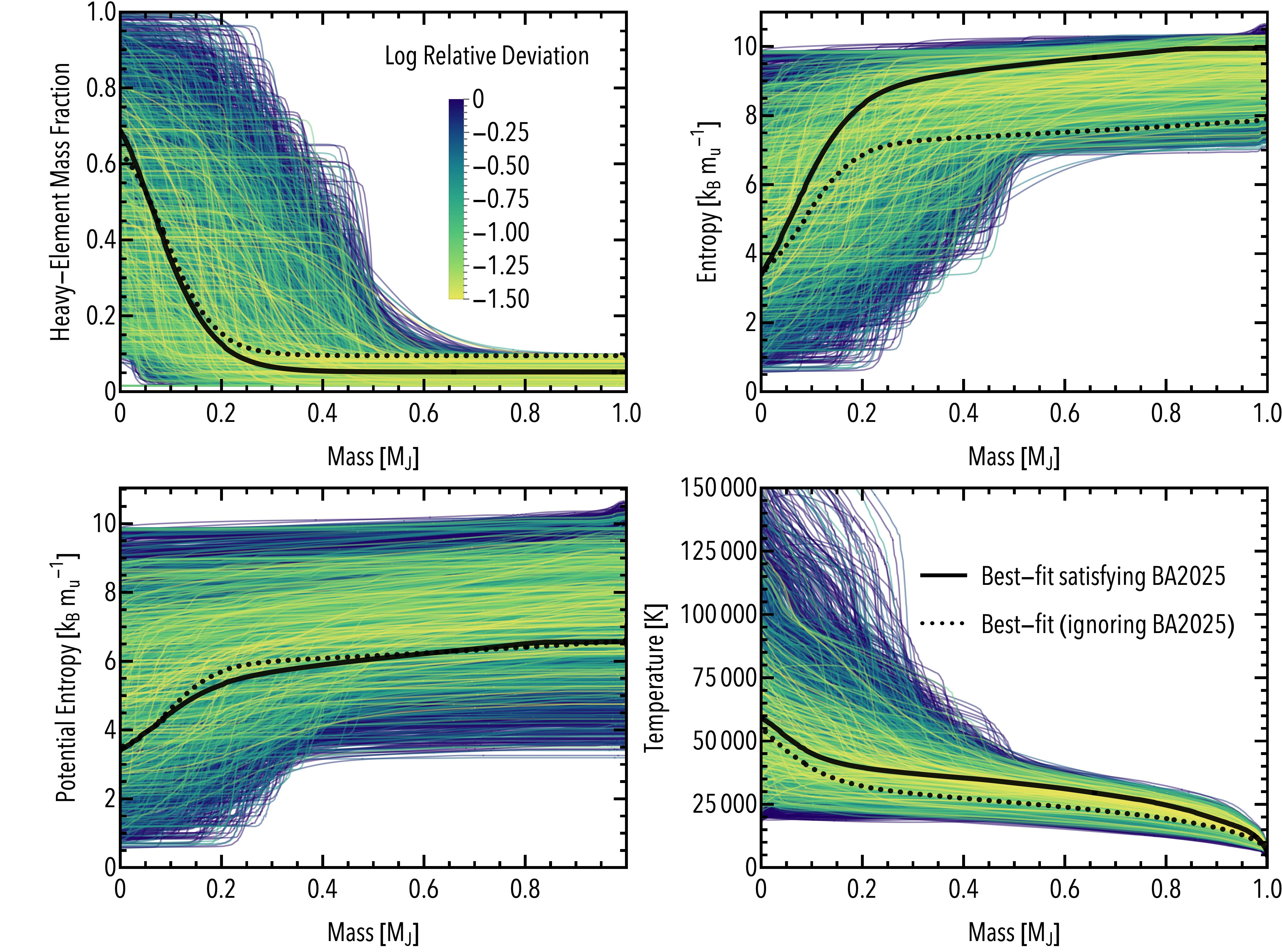}
\caption{Primordial heavy-element mass fraction (top left), specific entropy (top right), specific potential entropy (bottom left), and temperature (bottom right) as a function of mass for \nsims randomly generated Jupiter models.
The color scale indicates the quality of the fit to Jupiter's present-day structure. The black solid lines represent our best-fit model that fulfills Eq.~\eqref{eq:BA_2025}. The black dotted line represents the best-fit model without applying the BA2025 constraint.\label{fig:initial_profiles}}
\end{figure*}

\begin{figure*}[ht!]
\includegraphics[width=\textwidth]{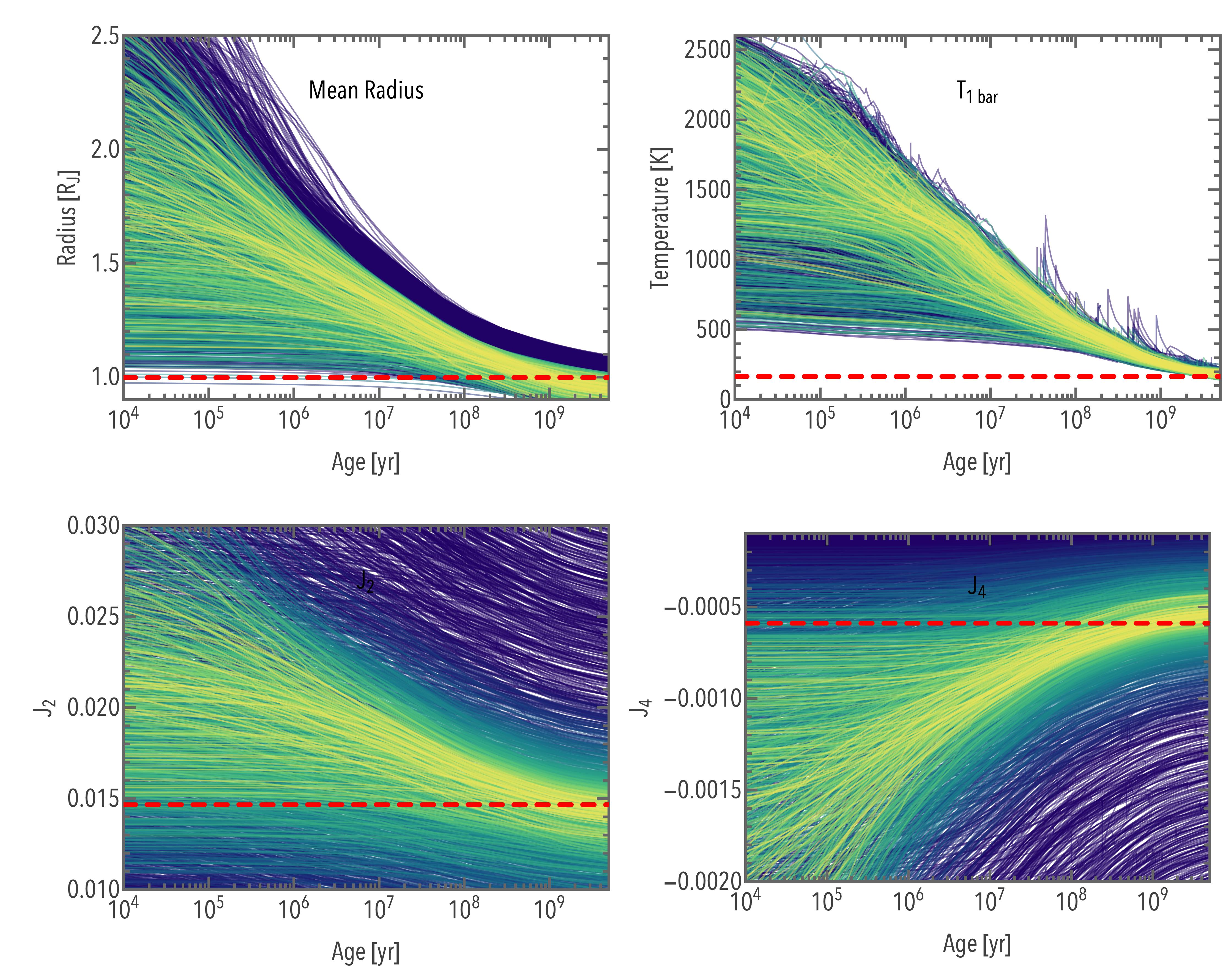}
\caption{Time evolution of the mean radius (top left), 1 bar temperature (top right), J$_2$ (bottom left), and J$_4$ (bottom right) compared to present-day Jupiter (dashed red line). The line colors are the same as in Fig. \ref{fig:initial_profiles}.\label{fig:evolution_overview}}
\end{figure*}

Figure~\ref{fig:initial_profiles} shows the initial conditions of our randomly generated planets and Fig.~\ref{fig:evolution_overview} shows their evolution over time.
Clearly, neither models with very shallow heavy-element distributions nor those with extended, high-Z cores are consistent with present-day Jupiter.
As expected, the best-fitting models share two key characteristics: (1) dilute cores extending to $\sim\SI{0.3}{\MJ}$, and (2) medium (specific) entropies in the core, typically around \SI{5}{\kbperbary} accounting for composition.
\par
The spikes observed in some of the 1-bar temperature profiles (Fig.~\ref{fig:evolution_overview}, top right) are caused by episodic mixing events. The primordial composition gradients in the dilute core models stabilize regions, effectively trapping interior heat. Core erosion triggers a rapid convective overturn, which efficiently releases the pent-up thermal energy. This sudden net release of energy---which dominates the gravitational cost of mixing the heavy elements---drives a sharp, temporary increase in the planet's luminosity and, consequently, its one-bar temperature.
\par
We also computed the specific potential entropy \citep[previously defined as effective entropy in][]{Knierim_2024}: 
\begin{equation}
    s_\mathrm{pot}(m) = s(m) - \sum_{i = 1}^{N-1} \int_{0}^{m} \diffp*{s}{{X_i}}{\,P, \rho, \{X_{j\neq i}\}} \diff{X_i}{m} \dint{m},
\end{equation}
where $s$ is the specific entropy, $m$ is the mass coordinate, $X_i$ is the mass fraction of chemical species $i$, and $N$ is the total number of species.
This quantity encapsulates both the specific entropy and the structural stability induced by the composition gradient.
As shown in Fig.~\ref{fig:initial_profiles}, and predicted in \citet{Knierim_2025}, the best-fitting models collapse onto a well-defined range of potential entropy profiles.
\par
\begin{figure}
\centering
\includegraphics[width=\columnwidth]{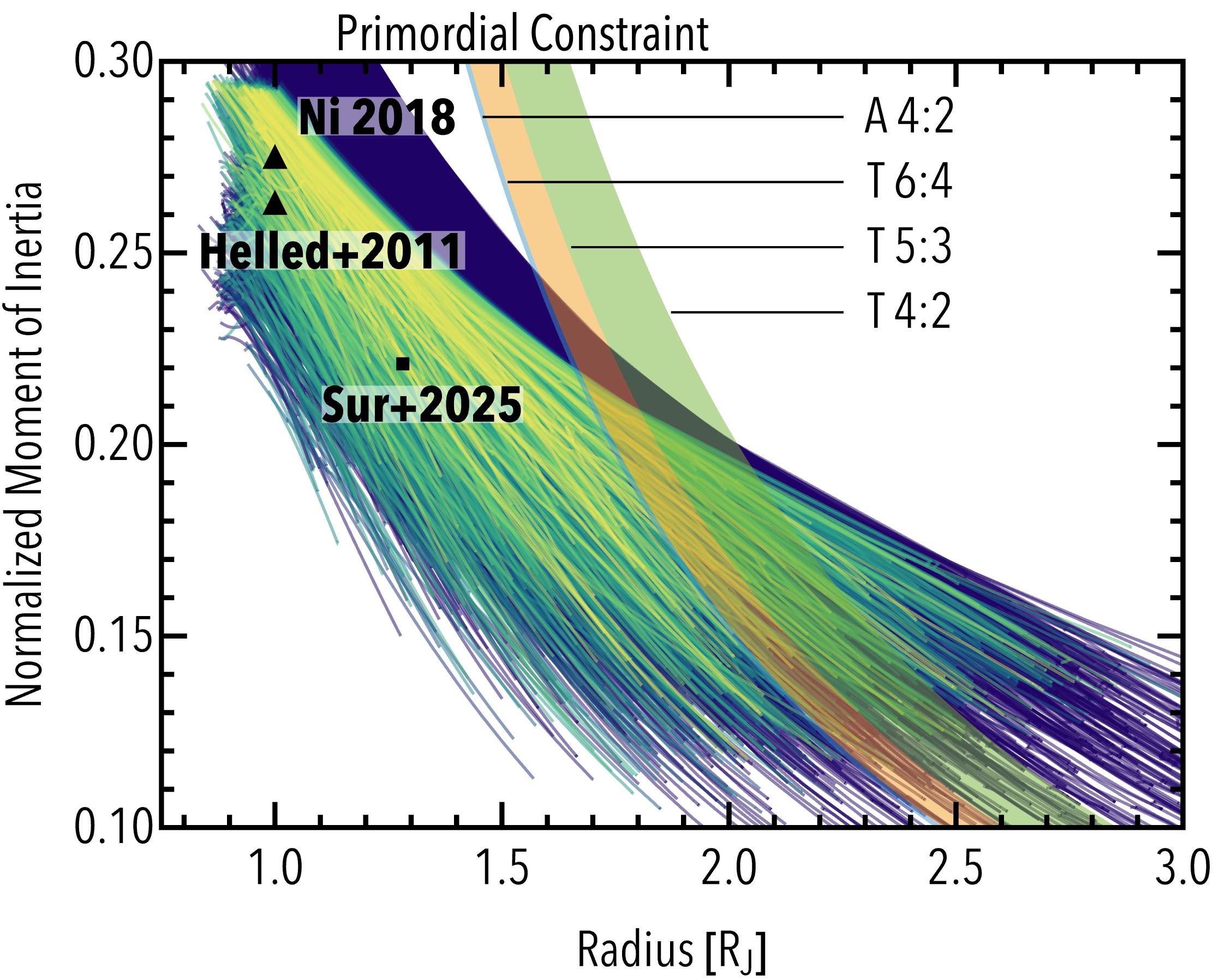}
\caption{Time evolution of normalized moment of inertia vs. radius for the sample of randomly generated Jupiter models. The line colors are the same as in Fig. \ref{fig:initial_profiles}. The colored regions indicate the primordial constraint of Eq.~\eqref{eq:BA_2025}, where the labels indicate the resonance of Io with Amalthea (A) or Thebe (T), respectively. The black triangles mark Jupiter's present-day radius and NMoI as estimated by \citet{Helled2011} and \citet{Ni2018}, respectively. The black square corresponds to the initial model from \citet{Sur2025}.\label{fig:NMOI_evolution}.}
\end{figure}

As Fig.~\ref{fig:NMOI_evolution} shows, not all of the well-fitting models are consistent with the primordial constraint from BA2025.
This discrepancy is driven almost entirely by the planet's initial entropy. The planetary entropy depends on both its thermal state and its composition. To isolate the thermal component from the effects of the composition gradient, we define a proto-solar reference entropy ($s_{\odot}$) as the entropy the structure would have if its composition were uniformly proto-solar. This approach allows for a direct comparison with previous studies that assume uniform composition, such as BA2025 or \citet[][]{Cumming2018}.

\begin{figure*}[ht!]
\includegraphics[width=\textwidth]{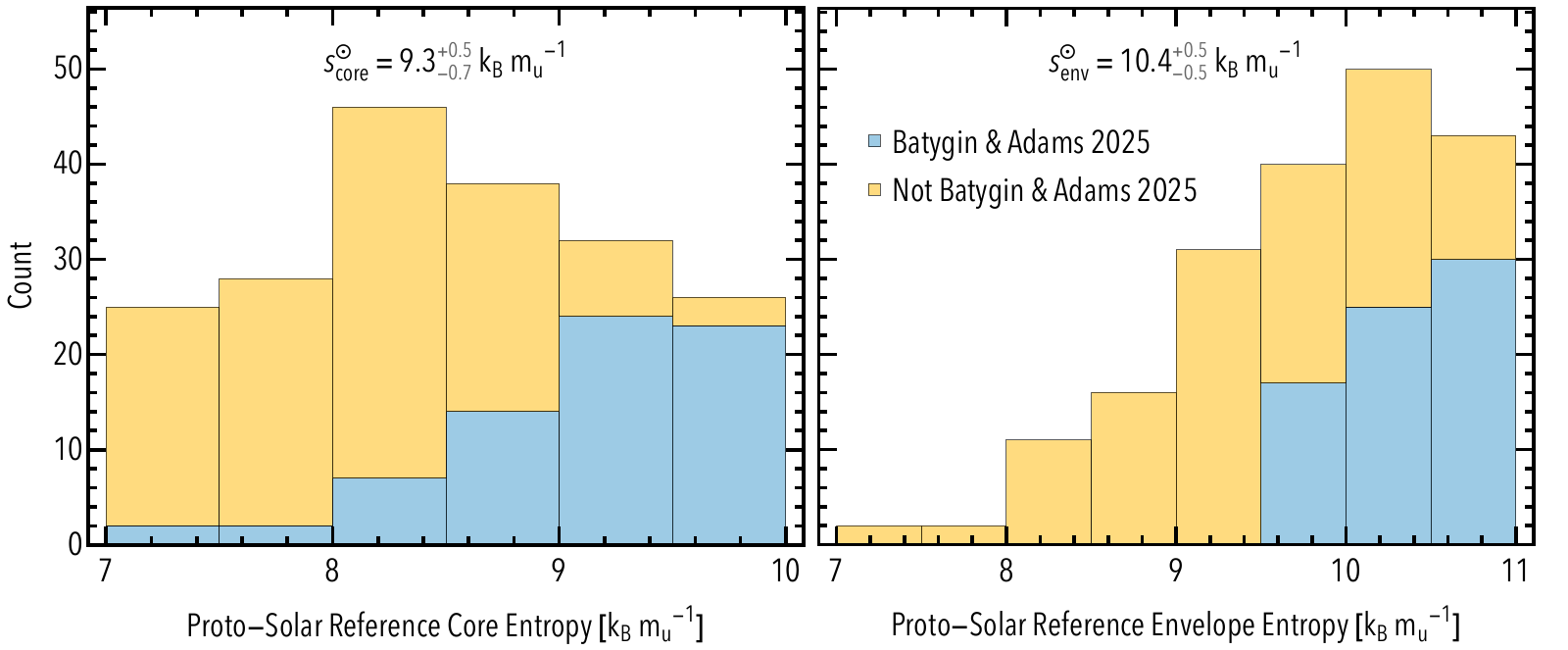}
\caption{Distributions of core entropy (left) and envelope entropy (right) before relaxing the composition gradient (i.e., at proto-solar composition) for models matching present-day Jupiter within \dHist. Colors of the stacked histogram indicate whether or not the primordial constraint of BA2025 is fulfilled. Best-fit values for the constrained models are marked at the center of each panel. \label{fig:entropy_histogram}}
\end{figure*}

With this distinction, a clear trend emerges. Figure~\ref{fig:entropy_histogram} shows that models initiated with a low reference entropy ($s_{\odot} \lesssim \SI{8}{\kbperbary}$) fail to satisfy the BA2025 criterion.
If Eq.~\eqref{eq:BA_2025} reflects a true dynamical constraint, these cold models are too compact to be primordial. 
Crucially, because the best-fitting models share approximately the same potential entropy profile (Fig.~\ref{fig:initial_profiles}), this finding demonstrates that the dynamical constraint effectively breaks the degeneracy between the thermal and compositional contributions to the planet's structure.
In fact, the BA2025 criterion constrains the primordial state to a core reference entropy of $\sRefCore = \sRefCoreBestFit$ and an envelope reference entropy of $\sRefEnv = \sRefEnvBestFit$, consistent with BA2025. Accounting for composition, the best-fitting values are $\sCore = \sCoreBestFit$ and $\sEnv = \sEnvBestFit$ for the core and envelope entropy, respectively. The resulting primordial radius is \RPrimBestFit, representing a slight downward correction compared to BA2025. For the influence of Eq.~\eqref{eq:BA_2025} on the initial primordial composition gradient, see Appendix \ref{sec:primordial_composition_gradient_constraints}. Thus, Eq.~\eqref{eq:BA_2025} further constrains Jupiter's primordial state.
\par
\begin{figure}[ht!]
\centering
\includegraphics[width=\columnwidth]{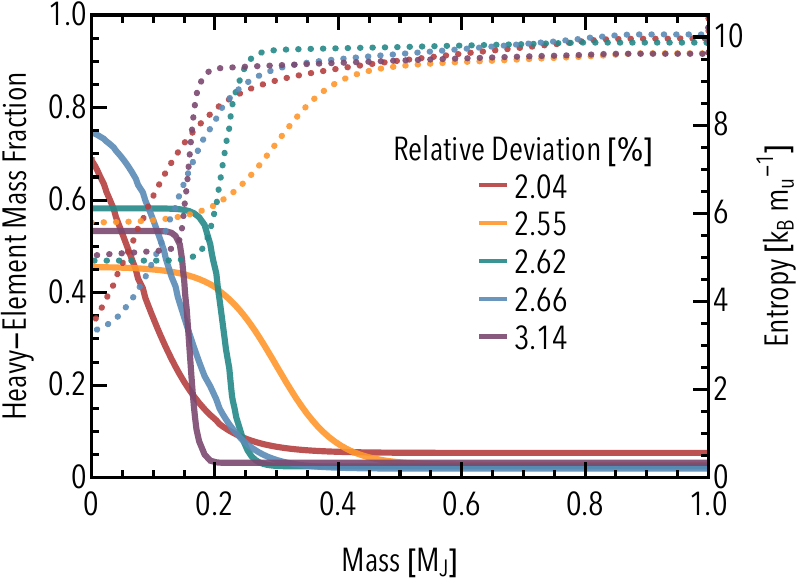}
\caption{Primordial heavy-element mass fraction (solid lines) and entropy (dotted lines) for the five best-fitting models that satisfy Eq.~\eqref{eq:BA_2025}.\label{fig:best_fitting_models}}
\end{figure}

Figure~\ref{fig:best_fitting_models} highlights the primordial structures of the five best-fitting models that satisfy Eq.~\eqref{eq:BA_2025}.
Notably, the heavy-element distributions are rather diverse. The central region could be a relatively compact step-like core that transitions near $\sim\SI{0.2}{\MJ}$ as expected from traditional formation models \citep{Mueller2020} or a highly extended ``fuzzy'' core with shallow gradients extending up to $\sim\SI{0.45}{\MJ}$, which would require a different formation history \citep{Helled2023}. We can therefore conclude that the primordial heavy-element distribution in Jupiter remains uncertain.
\par
Moreover, the emerging structure seems to be consistent with the core accretion scenario in which composition gradients are created during the planetary formation. In particular, it supports a formation path where most of the heavy elements are accreted during hydrostatic gas accretion (phase II) and subsequent runaway gas accretion (phase III) \citep[e.g.,][]{Shibata2023}, while phase III also delivers enough energy to produce a medium-to-high entropy envelope \citep[e.g.,][]{Cumming2018}.
\par
In addition to our sample, we included the evolution model of \citet{Sur2025} in Fig.~\ref{fig:NMOI_evolution}. While this model reproduces Jupiter's present-day structure remarkably well, it is inconsistent with Eq.~\eqref{eq:BA_2025}. The same applies to the model by \citet{Tejada_Arevalo2025}, which is very similar to the one from \citet{Sur2025} in Fig.~\ref{fig:NMOI_evolution} and was therefore omitted.
%

\section{Discussion} \label{sec:discussion}
The key uncertainties of this study are two-fold: Simplification in the numerical modeling and the reliability of Eq.~\eqref{eq:BA_2025}.
\subsection{Planetary evolution}\label{sec:discussion_planetary_evolution}
While planetary evolution models provide valuable insight into interior structure, they inevitably rely on approximations. In the following, we briefly discuss three key sources of uncertainty that may affect the quantitative details of our results.
\par
First, we model convection using mixing-length theory \citep[MLT; see][and references therein]{Kippenhahn2012}. In this framework, energy and material are primarily transported by the largest rising and sinking fluid parcels, which travels a characteristic mixing length $l_\mathrm{mlt} \propto \alpha H$, where $H$ is the pressure scale height and $\alpha$ is a free parameter typically of the order of unity. Material mixing is treated as a diffusive process, with a diffusion coefficient that also depends on $\alpha$.
Because convection is very efficient at removing both entropy and composition gradients, the details of MLT do not significantly affect the evolution within convective zones. However, whether a region becomes convective at all remains an open question, particularly in the context of oscillatory double-diffusive convection (ODDC), which can arise in layers that are Schwarzschild unstable but Ledoux stable \citep[e.g.,][]{Kato_1966,Radko_2007,Rosenblum_2011,Aurnou_2020,Fuentes_2020,Tulekeyev_2024}.
Our simulations do not model ODDC self-consistently. A more detailed treatment could therefore affect the structure and extent of Jupiter's fuzzy core.
\par
Second, our atmosphere model does not explicitly account for clouds or the detailed molecular composition of Jupiter's atmosphere \citep[e.g.,][]{Yi-Xian_2023}. While the observed conditions are approximated by adjusting the visible opacity in the semi-gray model of \citet{Guillot2010}, a fully self-consistent treatment would be preferable, as it reduces reliance on free parameters.
In planetary evolution models, the atmosphere primarily regulates the planetary contraction and cooling rate. This cooling drives the gradual expansion of the convective envelope, which in turn governs the erosion of the primordial composition profile \citep{Knierim_2024}.
A more realistic atmosphere model could constrain the cooling history more tightly, thereby narrowing the range of viable initial conditions. However, it would not affect the key conclusions of this study, which remain robust for reasonable atmospheric assumptions \citep[see also Sect. 4.5 in][]{Knierim_2024}.
\par
Third, the heavy elements are represented by an ideal mixture of water and rock. The assumption of ideal mixing itself introduces an uncertainty \citep{Darafeyeu2024}. In reality, Jupiter's interior could include different compositions and a more complex distribution of heavy elements, with varying compositions and depth-dependent abundances. Finally, the equations of state for heavy elements are also uncertain under planetary conditions, which can affect the inferred interior structure \citep[e.g.,][]{Howard_2025}.
\subsection{Fine-tuning Jupiter}
Within our sample, several models reproduce present-day Jupiter reasonably well, with relative differences below \dmin.
As discussed in Sect.~\ref{sec:discussion_planetary_evolution}, however, the detailed properties of an individual model remain sensitive to our specific assumptions.
In practice, such models could be tuned to achieve a marginally better or worse fit, but this fine-tuning would not yield deeper physical insight, it would merely reflect the limitations of the numerical framework.
Because the entire ensemble was computed under consistent assumptions, the robust outcome lies in the common trends that emerge across models rather than in any single realization.
Our results should therefore be viewed as constraints on Jupiter's interior in a general sense, not as a definitive, one-to-one prediction of its primordial structure.
\subsection{The dynamical constraint}
The main uncertainty in Eq.~\eqref{eq:BA_2025} lies in the assumption that Io's primordial location coincided with the truncation radius of Jupiter's circumplanetary disk (CPD). While this radius represents a natural barrier to inward migration, it is not necessarily impenetrable. In particular, sufficiently massive outer companions can push inner bodies past the disk edge via resonant interactions--a mechanism proposed for the innermost planets of the TRAPPIST-1 system \citep{Pichierri_2024}.
Applied to the Galilean system, this scenario would imply that Io and Europa were pushed inside the CPD cavity by their more massive companion, Ganymede. Lacking an outer neighbor of comparable mass, Ganymede itself would have stalled near the disk's truncation radius.
In the most favorable case for this scenario—where Io is located at $\xi = 4.11$—the three moons would be locked in their tightest possible configuration: a 4:2:1 mean-motion resonance. The truncation radius must then increase by a factor $4^{2/3}$, yielding $R_t = \SI{10.35}{\RJ}$.
Jupiter's corresponding primordial spin, given by $\Omega_\mathrm{J}^\dagger = \chi \sqrt{G M_\mathrm{J} / R_t^3}$ \citep{Batygin_2018}, would be reduced by a factor of 4 compared to the baseline scenario.
Angular momentum conservation requires that $I_\mathrm{J}^\dagger {R_\mathrm{J}^\dagger}^2$ increase by the same factor.
\par
However, as shown in Sect.~\ref{sec:results}, models with fuzzy cores naturally evolve toward Jupiter's present-day structure, and thus represent at least an intermediate evolutionary state. An increase in the core entropy or a modification of the composition gradient would disrupt this agreement with current constraints. An increase in the envelope entropy does not alter the final structure - it merely delays cooling and prolongs the evolutionary pathway \citep[see][]{Knierim_2024}.
Therefore, the only viable way to reconcile our best-fit models with the required fourfold increase in $I_\mathrm{J}^\dagger {R_\mathrm{J}^\dagger}^2$ is by inflating the radiative envelope.
Such an inflation reduces $I$, forcing $R$ to increase at minimum by a factor of 2, but more likely far beyond that. Numerical experiments show that even at $\si{\RJ}^\dagger \approx \SI{3}{\RJ}$, the Kelvin-Helmholtz timescale of proto-Jupiter would be of the order of \SI{5e4}{\yr}.
Achieving inflation beyond that state would require an extreme internal energy and would evolve on unrealistically short timescales, rendering the solution physically implausible.
Moreover, recent formation models of Amalthea and Thebe suggest that they were pushed in by Io \citep{Brunton2025a, Brunton2025b}. Hence, a proto-Jupiter larger than \SI{4}{\RJ} would have engulfed (and thus destroyed) these two moons.
We therefore conclude that the CPD truncation radius could not have been this far out, and that it is unlikely that Io migrated into Jupiter's magnetospheric cavity.
\section{Conclusions} \label{sec:conclusions}
We identified the range of primordial interior structures that evolve toward present-day Jupiter while simultaneously satisfying dynamical constraints from Jupiter's moon system—specifically, the inclinations of Amalthea and Thebe.
Our key conclusions are:
\begin{enumerate}
    \item The dynamical constraints from Amalthea and Thebe provide a valuable new handle on Jupiter's primordial structure, helping to break degeneracies inherent in thermal evolution models alone.
    \item Jupiter's present-day structure is best reproduced by an initially warm ($\sCore \sim \sCoreApprox$), dilute core ($\Zcore \sim \ZCoreApprox$), surrounded by a warm ($\sEnv \sim \sEnvApprox$), moderately enriched envelope ($\Zenv \sim \ZEnvApprox$).
    \item The dilute core of present-day Jupiter is most likely a primordial feature—shaped during formation, not by subsequent evolution.
\end{enumerate}
Overall, our results demonstrate the power of dynamical constraints to complement traditional structure modeling, and reinforce the view that Jupiter--and potentially gas giants more generally--is neither a homogeneous sphere of hydrogen and helium, nor a simple core+envelope system. Rather, it is a complex, stratified object whose structure encodes the fingerprints of its formation.
Fueled by upcoming space missions and unprecedented data, planetary science is poised to move from classification to explanation—to uncover not just what planets are, but also how they form and evolve.

\begin{acknowledgements}
We thank Ankan Sur, Roberto Tejada Arevalo, and Adam Burrows for kindly providing their Jupiter models. This work has been carried out within the framework of the National Centre of Competence in Research PlanetS supported by the Swiss National Science Foundation under grants \texttt{51NF40\_182901}, \texttt{51NF40\_205606}, and \texttt{215634}. FCA is supported in part by Grant No. 2508843 from the National Science Foundation (USA) and by the Leinweber Institute for Theoretical Physics at the University of Michigan. 
\end{acknowledgements}

\bibliographystyle{aa}
\bibliography{main}

@ARTICLE{Brunton2025b,
       author = {{Brunton}, Ian R. and {Batygin}, Konstantin},
        title = "{A Constraint on the Density of Jupiter's Moon Thebe from Primordial Dynamics}",
      journal = {\apjl},
         year = 2025,
        month = sep,
       volume = {990},
       number = {1},
          eid = {L11},
        pages = {L11},
          doi = {10.3847/2041-8213/adfa94},
archivePrefix = {arXiv},
       eprint = {2508.10109},
 primaryClass = {astro-ph.EP},
       adsurl = {https://ui.adsabs.harvard.edu/abs/2025ApJ...990L..11B}
}

@ARTICLE{Brunton2025a,
       author = {{Brunton}, Ian R. and {Batygin}, Konstantin},
        title = "{On the Origin and Dynamical Evolution of Jupiter's Moon Amalthea}",
      journal = {\apj},
         year = 2025,
        month = sep,
       volume = {991},
       number = {1},
          eid = {15},
        pages = {15},
          doi = {10.3847/1538-4357/adf432},
archivePrefix = {arXiv},
       eprint = {2507.17464},
 primaryClass = {astro-ph.EP},
       adsurl = {https://ui.adsabs.harvard.edu/abs/2025ApJ...991...15B}
}

@ARTICLE{Sur2025b,
       author = {{Sur}, Ankan and {Burrows}, Adam and {Tejada Arevalo}, Roberto},
        title = "{The Evolution of Jupiter and Saturn as a function of the Parameter R$_ρ$}",
      journal = {in press in \apj},
         year = 2025,
        month = jun,
          eid = {arXiv:2506.19041},
        pages = {arXiv:2506.19041},
          doi = {10.48550/arXiv.2506.19041},
archivePrefix = {arXiv},
       eprint = {2506.19041},
 primaryClass = {astro-ph.EP},
       adsurl = {https://ui.adsabs.harvard.edu/abs/2025arXiv250619041S}
}

@ARTICLE{Darafeyeu2024,
       author = {{Darafeyeu}, Valiantsin and {Rimle}, Stephanie and {Mazzola}, Guglielmo and {Helled}, Ravit},
        title = "{The Linear Mixing Approximation in Silica{\textendash}Water Mixtures at Planetary Conditions}",
      journal = {\apj},
         year = 2024,
        month = nov,
       volume = {975},
       number = {2},
          eid = {255},
        pages = {255},
          doi = {10.3847/1538-4357/ad7e29},
archivePrefix = {arXiv},
       eprint = {2409.14932},
 primaryClass = {astro-ph.EP},
       adsurl = {https://ui.adsabs.harvard.edu/abs/2024ApJ...975..255D}
}

@ARTICLE{Helled2011,
       author = {{Helled}, Ravit and {Anderson}, John D. and {Schubert}, Gerald and {Stevenson}, David J.},
        title = "{Jupiter{\textquoteright}s moment of inertia: A possible determination by Juno}",
      journal = {\icarus},
         year = 2011,
        month = dec,
       volume = {216},
       number = {2},
        pages = {440-448},
          doi = {10.1016/j.icarus.2011.09.016},
archivePrefix = {arXiv},
       eprint = {1109.1627},
 primaryClass = {astro-ph.EP},
       adsurl = {https://ui.adsabs.harvard.edu/abs/2011Icar..216..440H}
}

@ARTICLE{Ni2018,
       author = {{Ni}, Dongdong},
        title = "{Empirical models of Jupiter's interior from Juno data. Moment of inertia and tidal Love number k$_{2}$}",
      journal = {\aap},
         year = 2018,
        month = may,
       volume = {613},
          eid = {A32},
        pages = {A32},
          doi = {10.1051/0004-6361/201732183},
       adsurl = {https://ui.adsabs.harvard.edu/abs/2018A&A...613A..32N}
}

@software{PyToF,
  author       = {Morf, Luca},
  title        = {PyToF: a numerical implementation of the Theory of
                   Figures algorithm (4th, 7th, 10th order) including
                   barotropic differential rotation
                  },
  month        = aug,
  year         = 2025,
  publisher    = {Zenodo},
  version      = {v1.4.3},
  doi          = {10.5281/zenodo.16902935},
  url          = {https://doi.org/10.5281/zenodo.16902935},
}

@ARTICLE{Meier2025,
       author = {{Meier}, Thomas and {Reinhardt}, Christian and {Shibata}, Sho and {M{\"u}ller}, Simon and {Stadel}, Joachim and {Helled}, Ravit},
        title = "{On the Origin of Jupiter's Fuzzy Core: Constraints from N-body, Impact, and Evolution Simulations}",
      journal = {\apj},
         year = 2025,
        month = jul,
       volume = {988},
       number = {1},
          eid = {7},
        pages = {7},
          doi = {10.3847/1538-4357/addbe6},
archivePrefix = {arXiv},
       eprint = {2503.23997},
 primaryClass = {astro-ph.EP},
       adsurl = {https://ui.adsabs.harvard.edu/abs/2025ApJ...988....7M}
}

@ARTICLE{Nettelmann2025,
       author = {{Nettelmann}, N. and {Fortney}, J.~J.},
        title = "{Jupiter's Interior with an Inverted Helium Gradient}",
      journal = {\psj},
         year = 2025,
        month = apr,
       volume = {6},
       number = {4},
          eid = {98},
        pages = {98},
          doi = {10.3847/PSJ/adbdb7},
archivePrefix = {arXiv},
       eprint = {2504.00228},
 primaryClass = {astro-ph.EP},
       adsurl = {https://ui.adsabs.harvard.edu/abs/2025PSJ.....6...98N}
}

@ARTICLE{Pichierri_2024,
       author = {{Pichierri}, Gabriele and {Morbidelli}, Alessandro and {Batygin}, Konstantin and {Brasser}, Ramon},
        title = "{The formation of the TRAPPIST-1 system in two steps during the recession of the disk inner edge}",
      journal = {Nat. Astron.},
         year = 2024,
        month = nov,
       volume = {8},
        pages = {1408-1415},
          doi = {10.1038/s41550-024-02342-4},
archivePrefix = {arXiv},
       eprint = {2406.08677},
 primaryClass = {astro-ph.EP},
       adsurl = {https://ui.adsabs.harvard.edu/abs/2024NatAs...8.1408P}
}

@ARTICLE{Howard_2025,
       author = {{Howard}, S. and {Helled}, R. and {M{\"u}ller}, S.},
        title = "{Giant exoplanet composition: The impact of the hydrogen{\textendash}helium equation of state and interior structure}",
      journal = {\aap},
         year = 2025,
        month = jan,
       volume = {693},
          eid = {L7},
        pages = {L7},
          doi = {10.1051/0004-6361/202452783},
archivePrefix = {arXiv},
       eprint = {2410.21382},
 primaryClass = {astro-ph.EP},
       adsurl = {https://ui.adsabs.harvard.edu/abs/2025A&A...693L...7H}
}

@ARTICLE{Yi-Xian_2023,
       author = {{Chen}, Yi-Xian and {Burrows}, Adam and {Sur}, Ankan and {Arevalo}, Roberto Tejada},
        title = "{Jupiter Atmospheric Models and Outer Boundary Conditions for Giant Planet Evolutionary Calculations}",
      journal = {\apj},
         year = 2023,
        month = nov,
       volume = {957},
       number = {1},
          eid = {36},
        pages = {36},
          doi = {10.3847/1538-4357/acf456},
archivePrefix = {arXiv},
       eprint = {2309.00820},
 primaryClass = {astro-ph.EP},
       adsurl = {https://ui.adsabs.harvard.edu/abs/2023ApJ...957...36C}
}

@ARTICLE{Batygin_2018,
       author = {{Batygin}, Konstantin},
        title = "{On the Terminal Rotation Rates of Giant Planets}",
      journal = {\aj},
         year = 2018,
        month = apr,
       volume = {155},
       number = {4},
          eid = {178},
        pages = {178},
          doi = {10.3847/1538-3881/aab54e},
archivePrefix = {arXiv},
       eprint = {1803.07106},
 primaryClass = {astro-ph.EP},
       adsurl = {https://ui.adsabs.harvard.edu/abs/2018AJ....155..178B}
}

@INPROCEEDINGS{Hamilton_2001,
       author = {{Hamilton}, D.~P. and {Proctor}, A.~L. and {Rauch}, K.~P.},
        title = "{An Explanation for the High Inclinations of Thebe and Amalthea}",
    booktitle = {AAS/Division for Planetary Sciences Meeting Abstracts \#33},
         year = 2001,
       series = {AAS/Division for Planetary Sciences Meeting Abstracts},
       volume = {33},
        month = nov,
          eid = {25.04},
        pages = {25.04},
       adsurl = {https://ui.adsabs.harvard.edu/abs/2001DPS....33.2504H}
}

@ARTICLE{Gupta_2022,
       author = {{Gupta}, Pranika and {Atreya}, Sushil K. and {Steffes}, Paul G. and {Fletcher}, Leigh N. and {Guillot}, Tristan and {Allison}, Michael D. and {Bolton}, Scott J. and {Helled}, Ravit and {Levin}, Steven and {Li}, Cheng and {Lunine}, Jonathan I. and {Miguel}, Yamila and {Orton}, Glenn S. and {Hunter Waite}, J. and {Withers}, Paul},
        title = "{Jupiter's Temperature Structure: A Reassessment of the Voyager Radio Occultation Measurements}",
      journal = {\psj},
         year = 2022,
        month = jul,
       volume = {3},
       number = {7},
          eid = {159},
        pages = {159},
          doi = {10.3847/PSJ/ac6956},
archivePrefix = {arXiv},
       eprint = {2205.12926},
 primaryClass = {astro-ph.EP},
       adsurl = {https://ui.adsabs.harvard.edu/abs/2022PSJ.....3..159G}
}

@ARTICLE{Batygin2025,
       author = {{Batygin}, Konstantin and {Adams}, Fred C.},
        title = "{Determination of Jupiter's primordial physical state}",
      journal = {Nat. Astron.},
         year = 2025,
        month = jun,
       volume = {9},
        pages = {835-844},
          doi = {10.1038/s41550-025-02512-y},
archivePrefix = {arXiv},
       eprint = {2505.12652},
 primaryClass = {astro-ph.EP},
       adsurl = {https://ui.adsabs.harvard.edu/abs/2025NatAs...9..835B}
}

@ARTICLE{Tejada_Arevalo2025,
       author = {{Tejada Arevalo}, Roberto and {Sur}, Ankan and {Su}, Yubo and {Burrows}, Adam},
        title = "{Jupiter Evolutionary Models Incorporating Stably Stratified Regions}",
      journal = {\apj},
         year = 2025,
        month = feb,
       volume = {979},
       number = {2},
          eid = {243},
        pages = {243},
          doi = {10.3847/1538-4357/ada030},
       adsurl = {https://ui.adsabs.harvard.edu/abs/2025ApJ...979..243T}
}

@ARTICLE{Sur2025,
       author = {{Sur}, Ankan and {Tejada Arevalo}, Roberto and {Su}, Yubo and {Burrows}, Adam},
        title = "{Simultaneous Evolutionary Fits for Jupiter and Saturn Incorporating Fuzzy Cores}",
      journal = {\apjl},
         year = 2025,
        month = feb,
       volume = {980},
       number = {1},
          eid = {L5},
        pages = {L5},
          doi = {10.3847/2041-8213/adad62},
archivePrefix = {arXiv},
       eprint = {2412.17127},
 primaryClass = {astro-ph.EP},
       adsurl = {https://ui.adsabs.harvard.edu/abs/2025ApJ...980L...5S}
}

@ARTICLE{Zapolsky_1969,
       author = {{Zapolsky}, H.~S. and {Salpeter}, E.~E.},
        title = "{The Mass-Radius Relation for Cold Spheres of Low Mass}",
      journal = {\apj},
         year = 1969,
        month = nov,
       volume = {158},
        pages = {809},
          doi = {10.1086/150240},
       adsurl = {https://ui.adsabs.harvard.edu/abs/1969ApJ...158..809Z}
}

@ARTICLE{Kleine_2020,
       author = {{Kleine}, T. and {Budde}, G. and {Burkhardt}, C. and {Kruijer}, T.~S. and {Worsham}, E.~A. and {Morbidelli}, A. and {Nimmo}, F.},
        title = "{The Non-carbonaceous-Carbonaceous Meteorite Dichotomy}",
      journal = {\ssr},
         year = 2020,
        month = may,
       volume = {216},
       number = {4},
          eid = {55},
        pages = {55},
          doi = {10.1007/s11214-020-00675-w},
       adsurl = {https://ui.adsabs.harvard.edu/abs/2020SSRv..216...55K}
}

@ARTICLE{Batygin_2015,
       author = {{Batygin}, Konstantin and {Laughlin}, Greg},
        title = "{Jupiter's decisive role in the inner Solar System's early evolution}",
      journal = {Proc. Natl. Acad. Sci. U.S.A.},
         year = 2015,
        month = apr,
       volume = {112},
       number = {14},
        pages = {4214-4217},
          doi = {10.1073/pnas.1423252112},
archivePrefix = {arXiv},
       eprint = {1503.06945},
 primaryClass = {astro-ph.EP},
       adsurl = {https://ui.adsabs.harvard.edu/abs/2015PNAS..112.4214B}
}

@ARTICLE{Cozza_2025,
       author = {{Cozza}, Cesare and {Nakano}, Kousuke and {Howard}, Saburo and {Xie}, Hao and {Helled}, Ravit and {Mazzola}, Guglielmo},
        title = "{A Denser Hydrogen Inferred from First-Principles Simulations Challenges Jupiter's Interior Models}",
      journal = {under review at PRX},
         year = 2025,
        month = jan,
          eid = {arXiv:2501.12925},
        pages = {arXiv:2501.12925},
          doi = {10.48550/arXiv.2501.12925},
archivePrefix = {arXiv},
       eprint = {2501.12925},
 primaryClass = {astro-ph.EP},
       adsurl = {https://ui.adsabs.harvard.edu/abs/2025arXiv250112925C}
}

@ARTICLE{Knierim_2025,
       author = {{Knierim}, H. and {Helled}, R.},
        title = "{Unraveling the origin of giant exoplanets: Observational implications of convective mixing}",
      journal = {\aap},
         year = 2025,
        month = jun,
       volume = {698},
          eid = {L1},
        pages = {L1},
          doi = {10.1051/0004-6361/202554506},
archivePrefix = {arXiv},
       eprint = {2504.12118},
 primaryClass = {astro-ph.EP},
       adsurl = {https://ui.adsabs.harvard.edu/abs/2025A&A...698L...1K}
}

@ARTICLE{Knierim_2024,
       author = {{Knierim}, Henrik and {Helled}, Ravit},
        title = "{Convective Mixing in Gas Giant Planets with Primordial Composition Gradients}",
      journal = {\apj},
         year = 2024,
        month = dec,
       volume = {977},
       number = {2},
          eid = {227},
        pages = {227},
          doi = {10.3847/1538-4357/ad8dd0},
archivePrefix = {arXiv},
       eprint = {2407.09341},
 primaryClass = {astro-ph.EP},
       adsurl = {https://ui.adsabs.harvard.edu/abs/2024ApJ...977..227K}
}

@ARTICLE{Kato_1966,
       author = {{Kato}, S.},
        title = "{Overstable Convection in a Medium Stratified in Mean Molecular Weight}",
      journal = {\pasj},
         year = 1966,
        month = jan,
       volume = {18},
        pages = {374},
       adsurl = {https://ui.adsabs.harvard.edu/abs/1966PASJ...18..374K}
}

@ARTICLE{Tulekeyev_2024,
       author = {{Tulekeyev}, A. and {Garaud}, P. and {Idini}, B. and {Fortney}, J.~J.},
        title = "{Constraints on the Long-term Existence of Dilute Cores in Giant Planets}",
      journal = {\psj},
         year = 2024,
        month = aug,
       volume = {5},
       number = {8},
          eid = {190},
        pages = {190},
          doi = {10.3847/PSJ/ad6571},
archivePrefix = {arXiv},
       eprint = {2405.06790},
 primaryClass = {astro-ph.EP},
       adsurl = {https://ui.adsabs.harvard.edu/abs/2024PSJ.....5..190T}
}

@ARTICLE{Aurnou_2020,
       author = {{Aurnou}, Jonathan M. and {Horn}, Susanne and {Julien}, Keith},
        title = "{Connections between nonrotating, slowly rotating, and rapidly rotating turbulent convection transport scalings}",
      journal = {Phys. Rev. Research},
         year = 2020,
        month = oct,
       volume = {2},
       number = {4},
          eid = {043115},
        pages = {043115},
          doi = {10.1103/PhysRevResearch.2.043115},
archivePrefix = {arXiv},
       eprint = {2009.03447},
 primaryClass = {physics.flu-dyn},
       adsurl = {https://ui.adsabs.harvard.edu/abs/2020PhRvR...2d3115A}
}

@ARTICLE{Rosenblum_2011,
       author = {{Rosenblum}, E. and {Garaud}, P. and {Traxler}, A. and {Stellmach}, S.},
        title = "{Turbulent Mixing and Layer Formation in Double-diffusive Convection: Three-dimensional Numerical Simulations and Theory}",
      journal = {\apj},
         year = 2011,
        month = apr,
       volume = {731},
       number = {1},
          eid = {66},
        pages = {66},
          doi = {10.1088/0004-637X/731/1/66},
archivePrefix = {arXiv},
       eprint = {1012.0617},
 primaryClass = {astro-ph.SR},
       adsurl = {https://ui.adsabs.harvard.edu/abs/2011ApJ...731...66R}
}

@ARTICLE{Radko_2007,
       author = {{Radko}, Timour},
        title = "{Mechanics of merging events for a series of layers in a stratified turbulent fluid}",
      journal = {J. Fluid Mech.},
         year = 2007,
        month = apr,
       volume = {577},
        pages = {251},
          doi = {10.1017/S0022112007004703},
       adsurl = {https://ui.adsabs.harvard.edu/abs/2007JFM...577..251R}
}

@ARTICLE{Sur_2024,
       author = {{Sur}, Ankan and {Su}, Yubo and {Tejada Arevalo}, Roberto and {Chen}, Yi-Xian and {Burrows}, Adam},
        title = "{APPLE: An Evolution Code for Modeling Giant Planets}",
      journal = {\apj},
         year = 2024,
        month = aug,
       volume = {971},
       number = {1},
          eid = {104},
        pages = {104},
          doi = {10.3847/1538-4357/ad57c3},
archivePrefix = {arXiv},
       eprint = {2404.14483},
 primaryClass = {astro-ph.EP},
       adsurl = {https://ui.adsabs.harvard.edu/abs/2024ApJ...971..104S}
}

@ARTICLE{Fuentes_2020,
       author = {{Fuentes}, J.~R. and {Cumming}, A.},
        title = "{Penetration of a cooling convective layer into a stably-stratified composition gradient: Entrainment at low Prandtl number}",
      journal = {Phys. Rev. Fluids},
         year = 2020,
        month = dec,
       volume = {5},
       number = {12},
          eid = {124501},
        pages = {124501},
          doi = {10.1103/PhysRevFluids.5.124501},
archivePrefix = {arXiv},
       eprint = {2007.04265},
 primaryClass = {astro-ph.SR},
       adsurl = {https://ui.adsabs.harvard.edu/abs/2020PhRvF...5l4501F}
}

@ARTICLE{Stevenson2022,
       author = {{Stevenson}, David J. and {Bodenheimer}, Peter and {Lissauer}, Jack J. and {D'Angelo}, Gennaro},
        title = "{Mixing of Condensable Constituents with H-He during the Formation and Evolution of Jupiter}",
      journal = {\psj},
         year = 2022,
        month = apr,
       volume = {3},
       number = {4},
          eid = {74},
        pages = {74},
          doi = {10.3847/PSJ/ac5c44},
archivePrefix = {arXiv},
       eprint = {2202.09476},
 primaryClass = {astro-ph.EP},
       adsurl = {https://ui.adsabs.harvard.edu/abs/2022PSJ.....3...74S}
}

@ARTICLE{Bolton2017,
       author = {{Bolton}, S.~J. and {Lunine}, J. and {Stevenson}, D. and
         {Connerney}, J.~E.~P. and {Levin}, S. and {Owen}, T.~C. and
         {Bagenal}, F. and {Gautier}, D. and {Ingersoll}, A.~P. and
         {Orton}, G.~S. and {Guillot}, T. and {Hubbard}, W. and {Bloxham}, J. and
         {Coradini}, A. and {Stephens}, S.~K. and {Mokashi}, P. and
         {Thorne}, R. and {Thorpe}, R.},
        title = "{The Juno Mission}",
      journal = {\ssr},
         year = 2017,
        month = nov,
       volume = {213},
       number = {1-4},
        pages = {5-37},
          doi = {10.1007/s11214-017-0429-6},
       adsurl = {https://ui.adsabs.harvard.edu/abs/2017SSRv..213....5B}
}

@ARTICLE{Cumming2018,
       author = {{Cumming}, Andrew and {Helled}, Ravit and {Venturini}, Julia},
        title = "{The primordial entropy of Jupiter}",
      journal = {\mnras},
         year = 2018,
        month = jul,
       volume = {477},
       number = {4},
        pages = {4817-4823},
          doi = {10.1093/mnras/sty1000},
archivePrefix = {arXiv},
       eprint = {1804.06019},
 primaryClass = {astro-ph.EP},
       adsurl = {https://ui.adsabs.harvard.edu/abs/2018MNRAS.477.4817C}
}

@ARTICLE{Guillot2005,
       author = {{Guillot}, Tristan},
        title = "{THE INTERIORS OF GIANT PLANETS: Models and Outstanding Questions}",
      journal = {AREPS},
         year = 2005,
        month = jan,
       volume = {33},
        pages = {493-530},
          doi = {10.1146/annurev.earth.32.101802.120325},
archivePrefix = {arXiv},
       eprint = {astro-ph/0502068},
 primaryClass = {astro-ph},
       adsurl = {https://ui.adsabs.harvard.edu/abs/2005AREPS..33..493G}
}

@ARTICLE{Helled2017,
       author = {{Helled}, Ravit and {Stevenson}, David},
        title = "{The Fuzziness of Giant Planets{\textquoteright} Cores}",
      journal = {\apjl},
         year = 2017,
        month = may,
       volume = {840},
       number = {1},
          eid = {L4},
        pages = {L4},
          doi = {10.3847/2041-8213/aa6d08},
archivePrefix = {arXiv},
       eprint = {1704.01299},
 primaryClass = {astro-ph.EP},
       adsurl = {https://ui.adsabs.harvard.edu/abs/2017ApJ...840L...4H}
}

@BOOK{Kippenhahn2012,
       author = {{Kippenhahn}, Rudolf and {Weigert}, Alfred and {Weiss}, Achim},
        title = "{Stellar Structure and Evolution}",
        publisher = {Springer-Verlag Berlin Heidelberg},
         year = 2012,
          doi = {10.1007/978-3-642-30304-3},
       adsurl = {https://ui.adsabs.harvard.edu/abs/2012sse..book.....K}
}

@ARTICLE{Lozovsky2017,
       author = {{Lozovsky}, Michael and {Helled}, Ravit and {Rosenberg}, Eric D. and
         {Bodenheimer}, Peter},
        title = "{Jupiter{\textquoteright}s Formation and Its Primordial Internal Structure}",
      journal = {\apj},
         year = 2017,
        month = feb,
       volume = {836},
       number = {2},
          eid = {227},
        pages = {227},
          doi = {10.3847/1538-4357/836/2/227},
archivePrefix = {arXiv},
       eprint = {1701.01719},
 primaryClass = {astro-ph.EP},
       adsurl = {https://ui.adsabs.harvard.edu/abs/2017ApJ...836..227L}
}

@ARTICLE{Paxton2019,
       author = {{Paxton}, Bill and {Smolec}, R. and {Schwab}, Josiah and {Gautschy}, A. and
         {Bildsten}, Lars and {Cantiello}, Matteo and {Dotter}, Aaron and
         {Farmer}, R. and {Goldberg}, Jared A. and {Jermyn}, Adam S. and
         {Kanbur}, S.~M. and {Marchant}, Pablo and {Thoul}, Anne and
         {Townsend}, Richard H.~D. and {Wolf}, William M. and {Zhang}, Michael and
         {Timmes}, F.~X.},
        title = "{Modules for Experiments in Stellar Astrophysics (MESA): Pulsating Variable Stars, Rotation, Convective Boundaries, and Energy Conservation}",
      journal = {\apjs},
         year = 2019,
        month = jul,
       volume = {243},
       number = {1},
          eid = {10},
        pages = {10},
          doi = {10.3847/1538-4365/ab2241},
archivePrefix = {arXiv},
       eprint = {1903.01426},
 primaryClass = {astro-ph.SR},
       adsurl = {https://ui.adsabs.harvard.edu/abs/2019ApJS..243...10P}
}

@ARTICLE{Pollack1996,
       author = {{Pollack}, James B. and {Hubickyj}, Olenka and {Bodenheimer}, Peter and
         {Lissauer}, Jack J. and {Podolak}, Morris and {Greenzweig}, Yuval},
        title = "{Formation of the Giant Planets by Concurrent Accretion of Solids and Gas}",
      journal = {\icarus},
         year = 1996,
        month = nov,
       volume = {124},
       number = {1},
        pages = {62-85},
          doi = {10.1006/icar.1996.0190},
       adsurl = {https://ui.adsabs.harvard.edu/abs/1996Icar..124...62P}
}

@ARTICLE{Schoettler2018,
       author = {{Sch{\"o}ttler}, Manuel and {Redmer}, Ronald},
        title = "{Ab Initio Calculation of the Miscibility Diagram for Hydrogen-Helium Mixtures}",
      journal = {\prl},
         year = 2018,
        month = mar,
       volume = {120},
       number = {11},
          eid = {115703},
        pages = {115703},
          doi = {10.1103/PhysRevLett.120.115703},
       adsurl = {https://ui.adsabs.harvard.edu/abs/2018PhRvL.120k5703S}
}

@ARTICLE{Stevenson1985,
       author = {{Stevenson}, D.~J.},
        title = "{Cosmochemistry and structure of the giant planets and their satellites}",
      journal = {\icarus},
         year = 1985,
        month = apr,
       volume = {62},
       number = {1},
        pages = {4-15},
          doi = {10.1016/0019-1035(85)90168-X},
       adsurl = {https://ui.adsabs.harvard.edu/abs/1985Icar...62....4S}
}

@ARTICLE{Stevenson1982,
       author = {{Stevenson}, D.~J.},
        title = "{Interiors of the Giant Planets}",
      journal = {AREPS},
         year = 1982,
        month = jan,
       volume = {10},
        pages = {257},
          doi = {10.1146/annurev.ea.10.050182.001353},
       adsurl = {https://ui.adsabs.harvard.edu/abs/1982AREPS..10..257S}
}

@ARTICLE{Vazan2018,
       author = {{Vazan}, Allona and {Helled}, Ravit and {Guillot}, Tristan},
        title = "{Jupiter's evolution with primordial composition gradients}",
      journal = {\aap},
         year = 2018,
        month = feb,
       volume = {610},
          eid = {L14},
        pages = {L14},
          doi = {10.1051/0004-6361/201732522},
archivePrefix = {arXiv},
       eprint = {1801.08149},
 primaryClass = {astro-ph.EP},
       adsurl = {https://ui.adsabs.harvard.edu/abs/2018A&A...610L..14V}
}

@ARTICLE{Vazan2013,
       author = {{Vazan}, A. and {Kovetz}, A. and {Podolak}, M. and {Helled}, R.},
        title = "{The effect of composition on the evolution of giant and intermediate-mass planets}",
      journal = {\mnras},
         year = 2013,
        month = oct,
       volume = {434},
       number = {4},
        pages = {3283-3292},
          doi = {10.1093/mnras/stt1248},
archivePrefix = {arXiv},
       eprint = {1307.2033},
 primaryClass = {astro-ph.EP},
       adsurl = {https://ui.adsabs.harvard.edu/abs/2013MNRAS.434.3283V}
}

@ARTICLE{Wahl2017,
       author = {{Wahl}, S.~M. and {Hubbard}, W.~B. and {Militzer}, B. and {Guillot}, T. and
         {Miguel}, Y. and {Movshovitz}, N. and {Kaspi}, Y. and {Helled}, R. and
         {Reese}, D. and {Galanti}, E. and {Levin}, S. and {Connerney}, J.~E. and
         {Bolton}, S.~J.},
        title = "{Comparing Jupiter interior structure models to Juno gravity measurements and the role of a dilute core}",
      journal = {\grl},
         year = 2017,
        month = may,
       volume = {44},
       number = {10},
        pages = {4649-4659},
          doi = {10.1002/2017GL073160},
archivePrefix = {arXiv},
       eprint = {1707.01997},
 primaryClass = {astro-ph.EP},
       adsurl = {https://ui.adsabs.harvard.edu/abs/2017GeoRL..44.4649W}
}

@ARTICLE{Mueller2020,
       author = {{M{\"u}ller}, Simon and {Helled}, Ravit and {Cumming}, Andrew},
        title = "{The challenge of forming a fuzzy core in Jupiter}",
      journal = {\aap},
         year = 2020,
        month = jun,
       volume = {638},
          eid = {A121},
        pages = {A121},
          doi = {10.1051/0004-6361/201937376},
archivePrefix = {arXiv},
       eprint = {2004.13534},
 primaryClass = {astro-ph.EP},
       adsurl = {https://ui.adsabs.harvard.edu/abs/2020A&A...638A.121M}
}

@ARTICLE{Guillot2010,
       author = {{Guillot}, T.},
        title = "{On the radiative equilibrium of irradiated planetary atmospheres}",
      journal = {\aap},
         year = 2010,
        month = sep,
       volume = {520},
          eid = {A27},
        pages = {A27},
          doi = {10.1051/0004-6361/200913396},
archivePrefix = {arXiv},
       eprint = {1006.4702},
 primaryClass = {astro-ph.EP},
       adsurl = {https://ui.adsabs.harvard.edu/abs/2010A&A...520A..27G}
}

@ARTICLE{Chabrier_2021,
       author = {{Chabrier}, Gilles and {Debras}, Florian},
        title = "{A New Equation of State for Dense Hydrogen-Helium Mixtures. II. Taking into Account Hydrogen-Helium Interactions}",
      journal = {\apj},
         year = 2021,
        month = aug,
       volume = {917},
       number = {1},
          eid = {4},
        pages = {4},
          doi = {10.3847/1538-4357/abfc48},
archivePrefix = {arXiv},
       eprint = {2107.04434},
 primaryClass = {astro-ph.SR},
       adsurl = {https://ui.adsabs.harvard.edu/abs/2021ApJ...917....4C}
}

@ARTICLE{Shibata2023,
       author = {{Shibata}, S. and {Helled}, R. and {Kobayashi}, H.},
        title = "{Heavy-element accretion by proto-Jupiter in a massive planetesimal disc, revisited}",
      journal = {\mnras},
         year = 2023,
        month = feb,
       volume = {519},
       number = {2},
        pages = {1713-1731},
          doi = {10.1093/mnras/stac3568},
archivePrefix = {arXiv},
       eprint = {2212.04236},
 primaryClass = {astro-ph.EP},
       adsurl = {https://ui.adsabs.harvard.edu/abs/2023MNRAS.519.1713S}
}

@ARTICLE{Helled2023,
       author = {{Helled}, Ravit},
        title = "{The mass of gas giant planets: Is Saturn a failed gas giant?}",
      journal = {\aap},
         year = 2023,
        month = jul,
       volume = {675},
          eid = {L8},
        pages = {L8},
          doi = {10.1051/0004-6361/202346850},
archivePrefix = {arXiv},
       eprint = {2306.14740},
 primaryClass = {astro-ph.EP},
       adsurl = {https://ui.adsabs.harvard.edu/abs/2023A&A...675L...8H}
}

@ARTICLE{Helled2022_Jupiter,
       author = {{Helled}, Ravit and {Stevenson}, David J. and {Lunine}, Jonathan I. and {Bolton}, Scott J. and {Nettelmann}, Nadine and {Atreya}, Sushil and {Guillot}, Tristan and {Militzer}, Burkhard and {Miguel}, Yamila and {Hubbard}, William B.},
        title = "{Revelations on Jupiter's formation, evolution and interior: Challenges from Juno results}",
      journal = {\icarus},
         year = 2022,
        month = may,
       volume = {378},
          eid = {114937},
        pages = {114937},
          doi = {10.1016/j.icarus.2022.114937},
archivePrefix = {arXiv},
       eprint = {2202.10041},
 primaryClass = {astro-ph.EP},
       adsurl = {https://ui.adsabs.harvard.edu/abs/2022Icar..37814937H}
}

@ARTICLE{Stevenson_1977b,
       author = {{Stevenson}, D.~J. and {Salpeter}, E.~E.},
        title = "{The dynamics and helium distribution in hydrogen-helium fluid planets.}",
      journal = {\apjs},
         year = 1977,
        month = oct,
       volume = {35},
        pages = {239-261},
          doi = {10.1086/190479},
       adsurl = {https://ui.adsabs.harvard.edu/abs/1977ApJS...35..239S}
}

@ARTICLE{ni2019,
       author = {{Ni}, Dongdong},
        title = "{Understanding Jupiter's deep interior: the effect of a dilute core}",
      journal = {\aap},
         year = 2019,
        month = dec,
       volume = {632},
          eid = {A76},
        pages = {A76},
          doi = {10.1051/0004-6361/201935938},
       adsurl = {https://ui.adsabs.harvard.edu/abs/2019A&A...632A..76N}
}

@article{Jermyn_2023,
	author = {Jermyn, Adam S. and Bauer, Evan B. and Schwab, Josiah and Farmer, R. and Ball, Warrick H. and Bellinger, Earl P. and Dotter, Aaron and Joyce, Meridith and Marchant, Pablo and Mombarg, Joey S. G. and Wolf, William M. and Sunny Wong, Tin Long and Cinquegrana, Giulia C. and Farrell, Eoin and Smolec, R. and Thoul, Anne and Cantiello, Matteo and Herwig, Falk and Toloza, Odette and Bildsten, Lars and Townsend, Richard H. D. and Timmes, F. X.},
	doi = {10.3847/1538-4365/acae8d},
	issn = {1538-4365},
	journal = {ApJS},
	month = feb,
	number = {1},
	pages = {15},
	publisher = {American Astronomical Society},
	title = {Modules for Experiments in Stellar Astrophysics (MESA): Time-dependent Convection, Energy Conservation, Automatic Differentiation, and Infrastructure},
	url = {http://dx.doi.org/10.3847/1538-4365/acae8d},
	volume = {265},
	year = {2023}}

@article{Paxton_2018,
	author = {Paxton, Bill and Schwab, Josiah and Bauer, Evan B. and Bildsten, Lars and Blinnikov, Sergei and Duffell, Paul and Farmer, R. and Goldberg, Jared A. and Marchant, Pablo and Sorokina, Elena and Thoul, Anne and Townsend, Richard H. D. and Timmes, F. X.},
	doi = {10.3847/1538-4365/aaa5a8},
	issn = {1538-4365},
	journal = {ApJS},
	month = feb,
	number = {2},
	pages = {34},
	publisher = {American Astronomical Society},
	title = {Modules for Experiments in Stellar Astrophysics (${\mathtt{M}}{\mathtt{E}}{\mathtt{S}}{\mathtt{A}}$): Convective Boundaries, Element Diffusion, and Massive Star Explosions},
	url = {http://dx.doi.org/10.3847/1538-4365/aaa5a8},
	volume = {234},
	year = {2018}}

@article{Paxton_2015,
	author = {Paxton, Bill and Marchant, Pablo and Schwab, Josiah and Bauer, Evan B. and Bildsten, Lars and Cantiello, Matteo and Dessart, Luc and Farmer, R. and Hu, H. and Langer, N. and Townsend, R. H. D. and Townsley, Dean M. and Timmes, F. X.},
	doi = {10.1088/0067-0049/220/1/15},
	issn = {1538-4365},
	journal = {ApJS},
	month = sep,
	number = {1},
	pages = {15},
	publisher = {American Astronomical Society},
	title = {MODULES FOR EXPERIMENTS IN STELLAR ASTROPHYSICS (MESA): BINARIES, PULSATIONS, AND EXPLOSIONS},
	url = {http://dx.doi.org/10.1088/0067-0049/220/1/15},
	volume = {220},
	year = {2015}}

@article{Paxton_2013,
	author = {Paxton, Bill and Cantiello, Matteo and Arras, Phil and Bildsten, Lars and Brown, Edward F. and Dotter, Aaron and Mankovich, Christopher and Montgomery, M. H. and Stello, Dennis and Timmes, F. X. and Townsend, Richard},
	doi = {10.1088/0067-0049/208/1/4},
	issn = {1538-4365},
	journal = {ApJS},
	month = aug,
	number = {1},
	pages = {4},
	publisher = {American Astronomical Society},
	title = {MODULES FOR EXPERIMENTS IN STELLAR ASTROPHYSICS (MESA): PLANETS, OSCILLATIONS, ROTATION, AND MASSIVE STARS},
	url = {http://dx.doi.org/10.1088/0067-0049/208/1/4},
	volume = {208},
	year = {2013}}

@article{Paxton_2011,
	author = {Paxton, Bill and Bildsten, Lars and Dotter, Aaron and Herwig, Falk and Lesaffre, Pierre and Timmes, Frank},
	doi = {10.1088/0067-0049/192/1/3},
	issn = {1538-4365},
	journal = {ApJS},
	month = dec,
	number = {1},
	pages = {3},
	publisher = {American Astronomical Society},
	title = {MODULES FOR EXPERIMENTS IN STELLAR ASTROPHYSICS (MESA)},
	url = {http://dx.doi.org/10.1088/0067-0049/192/1/3},
	volume = {192},
	year = {2010}}

@BOOK{Zharkov_1978,
       author = {{Zharkov}, V.~N. and {Trubitsyn}, V.~P.},
    publisher = {Tucson, Ariz. : Pachart Pub. House},
        title = "{Physics of planetary interiors}",
         year = 1978,
       adsurl = {https://ui.adsabs.harvard.edu/abs/1978ppi..book.....Z}
}

\begin{appendix} 
\section{Rotation in \mesa}\label{sec:rotation}
In \mesa, rotation is implemented using the shellular approximation \citep{Paxton_2013}. Rather than solving the full three-dimensional structure, the object is divided into isobars, and the one-dimensional structure equations are solved as volume-averaged quantities over these isobars. Rotation enters explicitly into both the momentum balance and energy transport equations.

Angular momentum transport is treated as a diffusion process. While \mesa offers several prescriptions for differential rotation, for simplicity we assume solid-body rotation throughout this study.

Similarly, although \mesa includes models for rotationally induced mixing, magnetic field generation, or mass loss, these are primarily calibrated for stars and involve free parameters that are unknown for planets. We therefore omit these additional effects. Instead, we use the polar and equatorial radii, together with the rotation rate, to compute the gravitational moments and moment of inertia of the model.

Rotation is initialized using \texttt{relax\_initial\_omega}, which sets a uniform rotation rate throughout the planet at the start of our calculations.
\section{Random model generation}\label{sec:random_model_generation}
We construct the initial heavy-element distribution $Z(m)$ using a generalized logistic profile,
\begin{align}
Z(m) = \Zcore - \frac{\Zcore-\Zenv}{1+\exp\left[{\alpha_Z} (m - \mmid)\right]},
\end{align}
where $\Zcore$ and $\Zenv$ denote the heavy-element mass fractions in the limits $m \to -\infty$ and $m \to \infty$, respectively. The composition steepness parameter $\alpha_Z$ controls how steeply the heavy-element mass fraction transitions between these two regimes, and $\mmid$ specifies its midpoint in mass coordinate. For sufficiently large $\alpha_Z$, $\Zcore$ and $\Zenv$ may be interpreted as the heavy-element fractions at the center and in the outer envelope.
\par
The specific entropy profile at proto-solar composition, i.e., before relaxing the composition profile, is parameterized as
\begin{align}
s^{\odot}(m) = \sRefCore + \left(\frac{m}{M}\right)^{\alpha_s} \left(\sRefEnv - \sRefCore\right),
\end{align}
where the entropy steepness parameter $\alpha_s$ determines the slope between the core entropy $\sRefCore$ and the envelope entropy $\sRefEnv$. This form allows for a wide range of initial stratifications, from nearly isentropic interiors to steep entropy gradients, similar to those found in \citet{Cumming2018}.
\par
To explore a broad space of initial conditions, we draw parameters from wide ranges: $\Zcore$ uniformly between $0.1$ and $1$, $\Zenv$ uniformly between $Z_{\odot}$ and $0.1$, $\mmid$ uniformly between $0$ and \SI{0.5}{\MJ}, $\alpha_Z$ log-uniformly between $1$ and $200$, $\sRefCore$ uniformly between $7.0$ and \SI{10}{\kbperbary}, $\sRefEnv$ uniformly between $\sRefCore$ and \SI{11}{\kbperbary}, and $\alpha_s$ uniformly between $0.3$ and $2.0$. The total angular momentum is also varied by a factor of two around a reference value corresponding to a NMoI of $0.265$ \citep{Helled2011}.
Inside \mesa, we first relax the new angular momentum, then the entropy profile, and finally the composition profile before evolving the system.
\section{Constraints on the primordial composition gradient}\label{sec:primordial_composition_gradient_constraints}
\begin{figure*}[ht!]
\centering
\includegraphics[width=\textwidth]{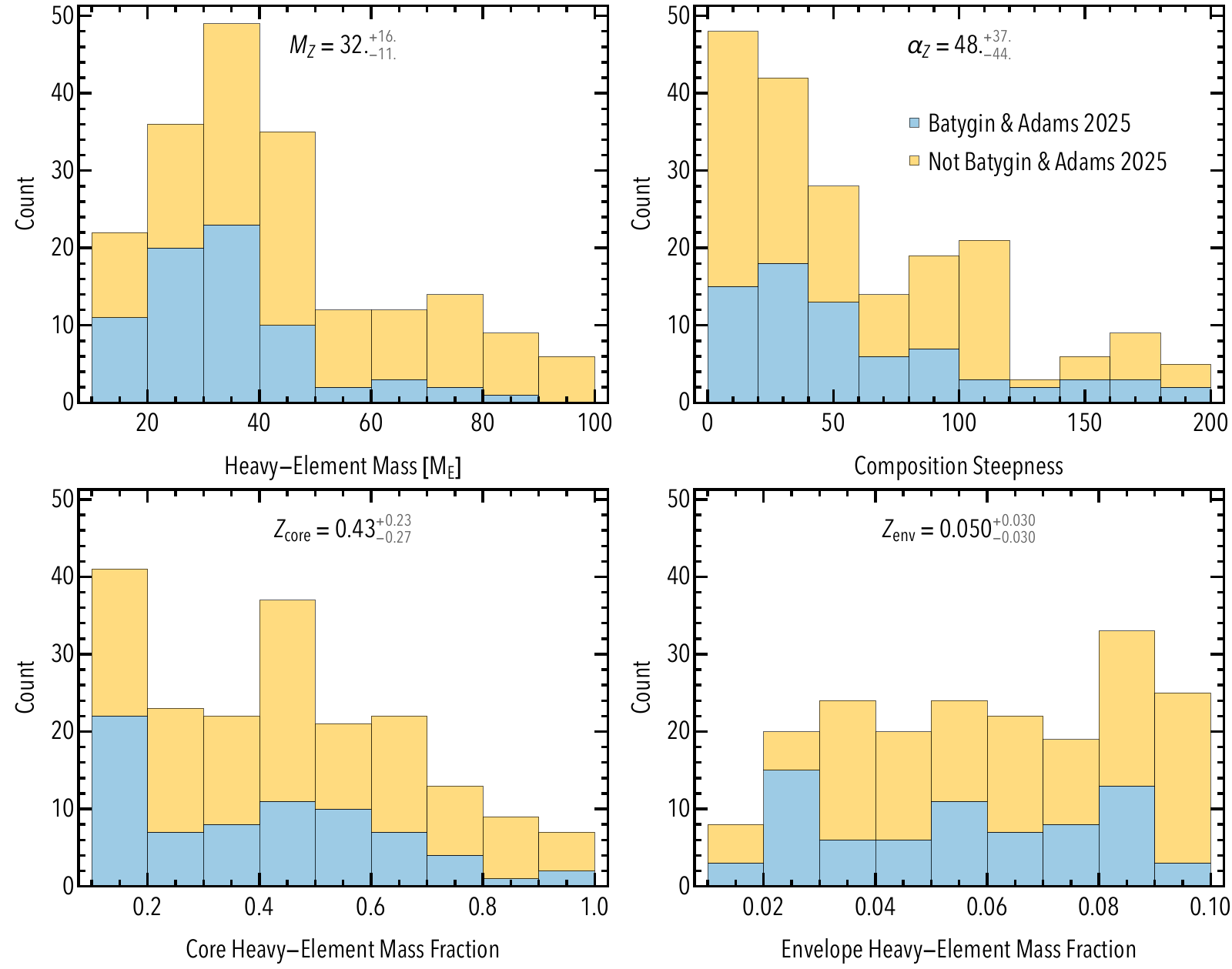}
\caption{Same as Fig. \ref{fig:entropy_histogram} but for the total heavy-element mass (top left), $\alpha_Z$ (top right), $\Zcore$ (bottom left), and $\Zenv$ (bottom right).\label{fig:primordial_gradient_constraints}}
\end{figure*}

Figure~\ref{fig:primordial_gradient_constraints} shows that, in contrast to the entropy results in Fig.~\ref{fig:entropy_histogram}, other primordial parameters are only weakly affected by the BA2025 constraint.
Models satisfying the BA2025 constraint tend to favor less massive cores, while showing no strong preference for the steepness of the composition profile. The distributions of core and envelope heavy-element fractions remain similar to those of the unconstrained case.
\end{appendix}

\end{document}